\documentclass{IEEEtran}
\usepackage[dvips]{graphicx}
\usepackage{epsfig,amssymb,amsmath,color,spconf, pstricks}
\usepackage{txfonts}

\newcommand{\beq}{\begin{equation}}
\newcommand{\eeq}{\end{equation}}
\newcommand{\beqn}{\begin{eqnarray}}

\newcommand{\eeqn}{\end{eqnarray}}

\title{Efficient Computation of Prolate Spheroidal Wave Functions in Radio Astronomical Source Modeling}

\twoauthors
{Parisa Noorishad}
{Kapteyn Astronomical Institute\\University of Groningen\\
Landleven 12, 9747 AV\\
Groningen, The Netherlands\\
Email: noorishad@astro.rug.nl}
{Sarod Yatawatta}
{ASTRON\\
Oude Hoogeveensedijk 4, 7991 PD\\
Dwingeloo, The Netherlands\\
Email: yatawatta@astron.nl}

\begin{document}

\maketitle
\begin{abstract}
The application of orthonormal basis functions such as Prolate Spheroidal Wave Functions (PSWF) for accurate source modeling in radio astronomy has been comprehensively studied. They are of great importance for high fidelity, high dynamic range imaging with new radio telescopes as well as conventional ones. But the construction of PSWF is computationally expensive compared to other closed form basis functions. In this paper, we suggest a solution to reduce its computational cost by more efficient construction of the matrix kernel which relates the image domain to visibility (or Fourier) domain. Radio astronomical images are mostly represented using a regular grid of rectangular pixels. This is required for efficient storage and display purposes and moreover, comes naturally as a by product of the Fast Fourier Transform (FFT) in imaging.  We propose the use of Delaunay triangulation as opposed to regular gridding of an image for a finer selection of the region of interest (signal support) during the PSWF kernel construction. We show that the computational efficiency improves without loss of information. Once the PSWF basis is constructed using the irregular grid, we revert back to the regular grid by interpolation and thereafter, conventional imaging techniques can be applied.
\end{abstract}
%\begin{keywords}

\section{Introduction}
For radio interferometric imaging, we measure the spatial coherency or Fourier components of the celestial sources on different baselines in our array (also called as visibilities). Due to a finite number of receivers, we only sample the Fourier components at a finite number of positions. After calibration or correction for the corruptions due to the propagation path and instrumental effects, we have to deconvolve the source flux from the antennae response pattern to obtain its true intensity map or its radio image.

Conventionally, CLEAN  \cite{Hogbom},\cite{Schwarz1978} algorithm has been used for modeling and removal of the known sources such that the faint unknown sources in the background can appear. It is a pixel based deconvolution approach. This means any point in the image is associated with a weighted delta function. CLEAN will fail to perform satisfactorily, if two point sources are too close or if the source is partially resolved, as it has analytically been proven in \cite{SAM2010}. Moreover, we have shown that making image grid smaller (expecting that accuracy of modeling extended sources improves), is a futile effort by presenting the Cramer Rao Bound (CRB) analysis \cite{SAM2010}. Instead, we showed that using a 2-D orthonormal basis for deconvolution achieves the theoretical noise bound or CRB, regardless of the source structure \cite{SAM2010}. 

Cartesian shapelets have been used for source modeling in \cite{SAM2010}, (and references therein). It has been proven that shapelets provide a way to model extended sources by a closed form construction of an orthonormal basis. But this technique requires additional constraints such as information theoretic bounds to limit the model order. In \cite{SYURSI2011}, \cite{ICIP2011}, we have shown that we can construct the optimal basis to model the source by minimum number of basis functions and less artifacts outside Region Of Interest (ROI) using prolate spheroidal wave functions. In addition, unlike shapelets, they do not require additional explicit constrains from information theory, as it is implicitly considered in their mathematical derivation (see \cite{ICIP2011}). However, the construction of PSWF is computationally more expensive as compared to shapelets \cite{SYURSI2011}. 

In this paper, we present computationally efficient construction of PSWF suitable for modeling extended sources. We achieve this by downsampling both the Fourier plane visibilities as well as image pixels. Most astronomical (as well as other) images are represented by a regular 2-D array of rectangular pixels. In contrast, we propose to use Delaunay triangulation to construct an irregular grid with fewer number of pixels outside ROI and finer selection of the image inside ROI to represent the image during the construction of PSWF. We can control the size of the triangulation without losing information. Once the PSWF basis has been constructed in this manner, we revert back to the original pixel grid by interpolation.

Notation: We denote vectors in bold lowercase and matrices in bold uppercase. The matrix transpose, Hermitian, and pseudoinverse are denoted by $(.)^T$ , $(.)^H$ and $(.)^{\dagger}$ respectively. Operation $\Arrowvert . \Arrowvert$ gives the Frobenious norm of a matrix.

\section{Mathematical Preliminaries}

\subsection{Radio interferometric imaging}

\begin{figure}[htb]
\begin{center}
\scalebox{1.25}{
 	\input{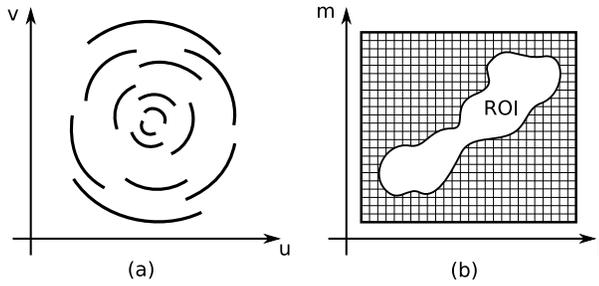}
	}
\caption{(a) Sampling points (total $N_a$) in the Fourier (visibility) plane. (b) Image of $N$ rectangular pixels, with the support (ROI) area in white. The ROI area has $N_b$ pixels.\label{fig:roi}}
\end{center}
\end{figure}

In radio interferometric imaging (see Fig. \ref{fig:roi}), the relation between the intensity of the $q$-th pixel, $f(l_q, m_q)$ in an image and the $p$-th point, $\widetilde{f}(u_p, v_p)$ in the visibility domain is described by van Cittert-Zernike theorem \cite{Brouw1975} as:

\beqn \label{eq:no.1}
f(l_q, m_q)=\sum_{p=0}^{N_a-1} \tilde{f}(u_p, v_p)e^{j2\pi(l_q u_p+ m_q v_p)}, \\\nonumber
\widetilde{f}(u_p, v_p) = \sum_{q=0}^{N-1} {f}(l_q, m_q)e^{-j2\pi(l_q u_p+ m_q v_p)},\\\nonumber
p\in{[0, N_a-1]},\ q\in{[0, N]},
\eeqn

\noindent where $N_a$ is the total sampling points in visibility domain. $N$ is the total number of pixels in the image. If we rewrite them in vectorized form, we will obtain:

%\begin{equation}
%\begin{eqnarray}
\begin{align}
&\widetilde{\mathbf{f}} = \mathbf{T}\mathbf{f},~~~ \mathbf{f}= \mathbf{T}^H\widetilde{\mathbf{f}}\\
	          &\mathbf{f} \equiv [f(l_0, m_0), ..., f(l_{N-1}, m_{N-1})]^H \nonumber \\
                  &\widetilde{\mathbf{f}} \equiv [\tilde{f}(u_0, v_0), ..., \widetilde{f}(u_{N_a-1}, v_{N_a-1})]^H. \nonumber
\label{eq:no.3}
%\triangleeq
%\end{eqnarray}
\end{align}
%\end{equation}

The matrix $\mathbf{T}$ (size $N_a\times N$) contains $e^{-j2\pi(l_q u_p+ m_q v_p)}$ on its $q$-th row and $p$-th column.

\subsection{Prolate spheroidal basis}

For a complete mathematical derivation of the PSWF, the reader is referred to \cite{ICIP2011} and references therein. In the following, we will mention a few key mathematical relations of PSWF derivation for stating our problem.

We aim to obtain representative basis, $p(l, m)$ for a source that maximizes the energy in our ROI i.e. within the boundary by which a source is recognized in an image. The larger the ratio given in (\ref{eq:no.4}), the less artifacts  will remain outside the ROI:

\begin{equation}
\lambda = \frac{\sum_{(l, m)\in{ROI}} \arrowvert{p(l, m)}\arrowvert^2}{\sum_{(l, m)} \arrowvert{p(l, m)}\arrowvert^2}
\label{eq:no.4}
\end{equation}

\noindent The vectorized form of which can be written as:

\begin{equation}
\lambda = \frac{\Arrowvert \mathbf{I}_b^T \mathbf{p}\Arrowvert^2}{\Arrowvert \mathbf{p} \Arrowvert^2}
\label{eq:no.5}
\end{equation}

\noindent where $\mathbf{I}_b^T$, (size $N_b\times N$) is a selection matrix to select the pixels that belong to the ROI. $N_b$ is the number of pixels in the ROI. Then the PSWF basis can be obtained by eigen-decomposition of the kernel, $\mathbf{K}$ (size $N_b\times N_b$) given as:

\begin{equation}
\mathbf{K} = \mathbf{I}_b^T \mathbf{T}^H (\mathbf{T}\mathbf{T}^H)^{\dagger} \mathbf{T} \mathbf{I}_b 
\label{eq:no.6}
\end{equation}

Most of the computations are required first to find the  pseudo-inverse term, $(\mathbf{T}\mathbf{T}^H)^{\dagger}$ (size $N_a\times N_a$) and  secondly, to find the eigenvalue decomposition of ${\bf K}$ (size $N_b\times N_b$). Note that the rank of $(\mathbf{T}\mathbf{T}^H)^{\dagger}$ is at most $N$ and usually $N_a\gg N> N_b$. By properly downsampling both the Fourier plane sampling points (reducing $N_a$) as well as the image pixels (reducing $N$ and $N_b$) we can significantly cut down the cost of computations. In the next section, we will explain how it has efficiently been done, followed by an example.

\section{Computational cost reduction}
We take several steps to reduce the computational cost of (\ref{eq:no.6}). The main criterion for this reduction is given by the Landau-Pollak theorem \cite{LandauPollak1961}. In its simplest form (applicable to radio interferometry) \cite{SYURSI2011}, Landau-Pollak theorem states that the number of degrees of freedom of any given source with compact support, $N_D$ could be no greater than the product of area of the source, $A_{lm}$ and the area of the source observed in the Fourier plane, $A_{uv}$ i.e. $N_D \leq A_{lm} \times A_{uv}$. 

\subsection{Reducing $N_a$}
Since we already have an irregular set of $N_a$ sampling points on the Fourier plane, it is straightforward to downsample this to a lower value $N_a^{'}$, which is already presented in \cite{ICIP2011}. However, we can do even better by combining this with imaging weights. Each point on the Fourier plane, (say $p$) has a weight associated with it, $\rho(u_p,v_p) \in [0,1]$.  Prior to taking the Fourier transform in (\ref{eq:no.1}), we multiply each data point $\widetilde{f}(u_p,v_p)$ by this weight. When we downsample, we generate a uniformly distributed random number $r(u_p,v_p) \in [0,1]$ and if $r(u_p,v_p) \le \rho(u_p,v_p)$ we select the $p$-th point.

\subsection{Reducing $N$ but finer selection of ROI}
\begin{figure}[htb]
\begin{center}
\scalebox{1.2}{
 	\input{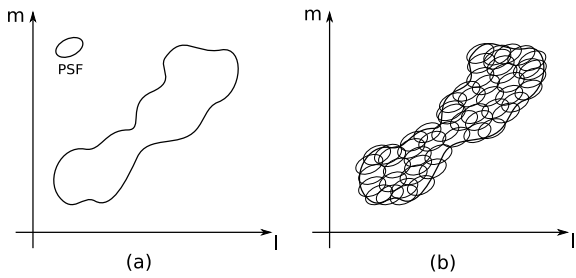}
	}
\caption{(a) The ROI of an extended source and the PSF (the ellipse). (b) The ROI covered by the PSF (an ellipse packing).\label{fig:cp}}
\end{center}
\end{figure}

The motivation behind reducing the number of image pixels is that not all the pixels are required to define the ROI (because of Landau-Pollak theorem) while we construct the PSWF. In Fig. \ref{fig:cp} (a), we have shown the ROI of the source as well as the Point Spread Function (PSF). The amount of information can be loosely translated to the number of PSFs that we can pack within the ROI, as shown in Fig. \ref{fig:cp} (b). The optimal way to pack the PSFs within the ROI boils down to a circle packing problem, which is generally understood as NP-hard.

\begin{figure}[htbp]
\begin{minipage}{0.98\linewidth}
\centering
 \centerline{\epsfig{figure=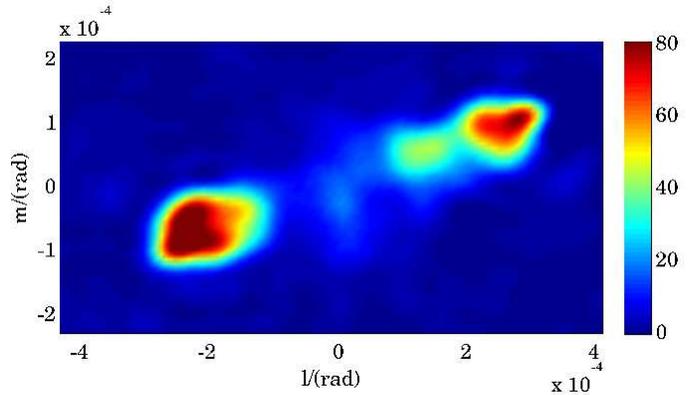,width=9.0cm}}
\end{minipage}
\caption{A test image of an extended radio source, with dimensions $118\times64$  pixels.\label{image1}}
\end{figure}

Therefore, we follow a heuristic approach: Instead of solving a packing problem, we construct an irregular grid of triangles to cover the ROI as well as the area of the image outside the ROI. Obviously, we employ Delaunay triangulation \cite{DT} for this purpose. We shall illustrate this with an example as follows: In Fig. \ref{image1}, we have shown an image of an extended radio source. This image (dimensions $118\times64$) has $7552$ pixels.

\begin{figure}[htbp]
\begin{minipage}{0.98\linewidth}
\centering
 \centerline{\epsfig{figure=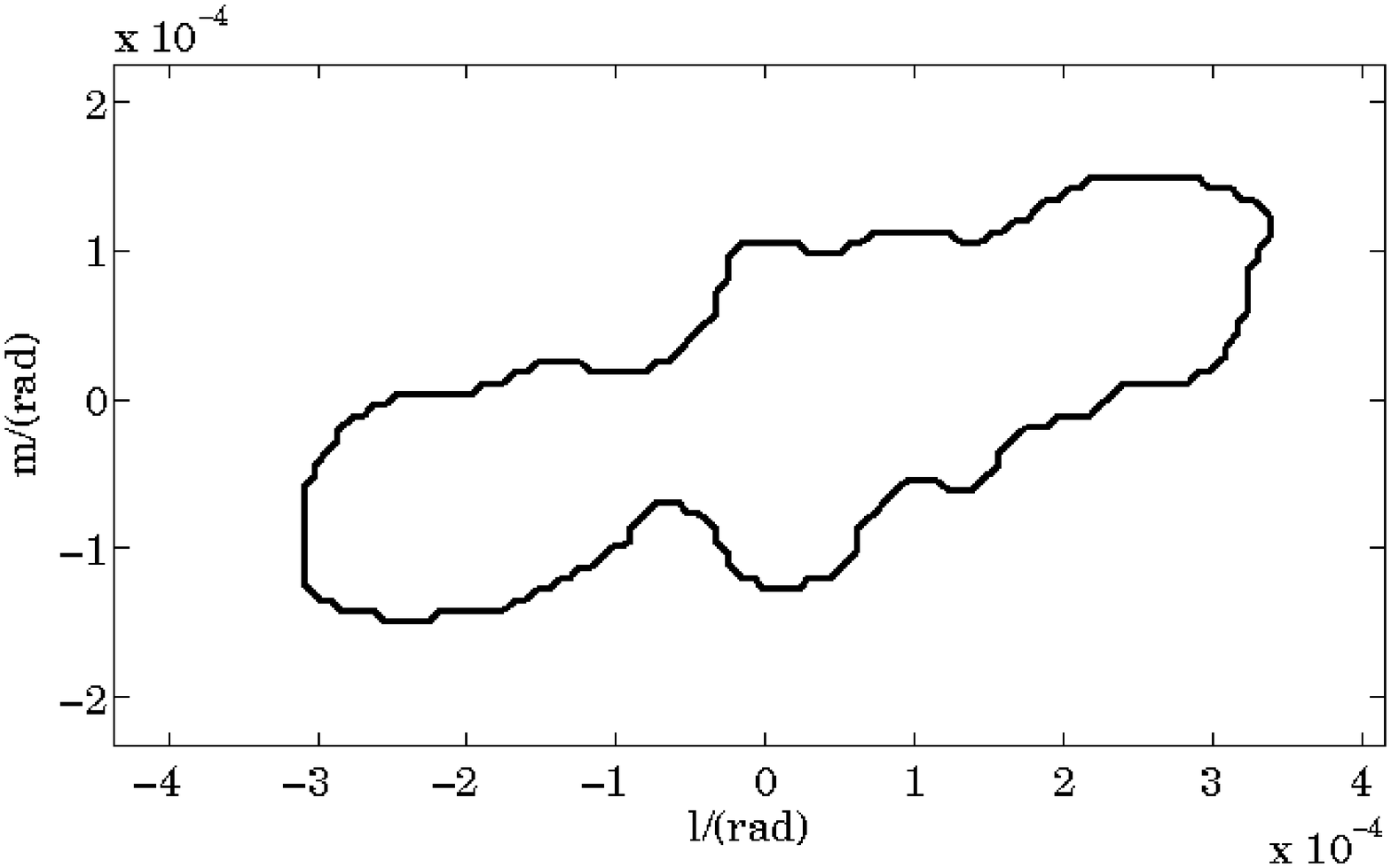,width=9.0cm}}
\end{minipage}
\caption{The ROI of the source in Fig. \ref{image1}. The total image has $7552$ pixels while the ROI has $1826$ pixels.\label{roi}}
\end{figure}

\begin{figure}[htbp]
\begin{minipage}{0.98\linewidth}
\centering
 \centerline{\epsfig{figure=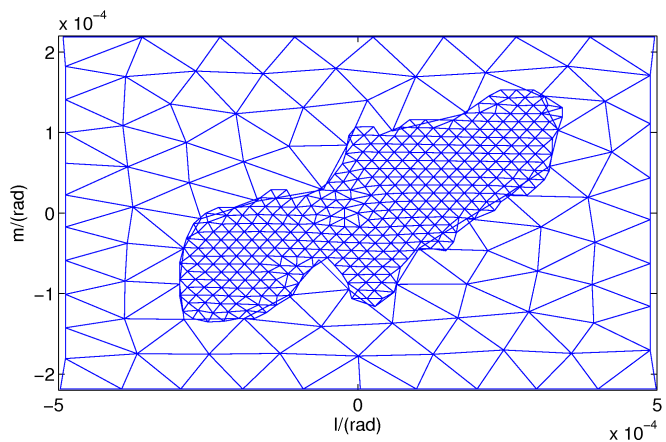,width=9.0cm}}
\end{minipage}
\caption{Delaunay triangulation of the image in Fig. \ref{image1}. The triangulation inside the ROI has $1080$ triangles while the triangulation outside the ROI has $140$ triangles. The triangle size within the ROI is made smaller than the exterior triangles.\label{tri}}
\end{figure}

In Fig. \ref{roi}, we have shown the ROI for this source, which has $1826$ pixels to represent the amount of information present. Therefore, we have generated the Delaunay triangulation for the ROI and its exterior as shown in Fig. \ref{tri} such that the triangle size within the ROI is smaller than the exterior triangles. We take the centroids of the triangles as our reduced grid to generate the PSWF. Therefore, we end up with a total number of pixels of $1220$ while the ROI has $1080$ pixels.

\subsection{Reducing cost of pseudoinverse}
The cost of (\ref{eq:no.6}) mostly involves evaluation of the pseudoinverse $(\mathbf{T}\mathbf{T}^H)^{\dagger}$. Instead of directly evaluating this, we solve a set of linear equations such as
\begin{equation}
(\mathbf{T}\mathbf{T}^H + \gamma {\bf I}) {\bf x}  = {\bf A} {\bf x}= {\bf y} 
\label{eq:no7}
\end{equation}
where $\gamma$ is a small positive regularization parameter. Depending on the context, we are given ${\bf y}$ and we need to find ${\bf x}$. The solution of the linear system for ${\bf x}$ is done using Q-less QR factorization.

Given ${\bf Q}{\bf R}={\bf A}$ (where ${\bf Q}$ is unitary and ${\bf R}$ is upper triangular), we first solve
\begin{equation}
{\bf R}^H{\bf R}\widetilde{\bf x}={\bf A}^H{\bf y}
\label{eq:no.8}
\end{equation}
\noindent  for $\widetilde{\bf x}$ and make one step correction

\beqn\label{eq:no.9}
{\bf r}={\bf y}-{\bf A}\widetilde{\bf x}\\\nonumber
{\mathrm solve}\ \ {\bf R}^H{\bf R}\widetilde{\bf e}={\bf A}^H{\bf r}\\\nonumber
{\bf x}=\widetilde{\bf x}+ \widetilde{\bf e}
\eeqn
\noindent to get the solution ${\bf x}$.

\section{Results}
To demonstrate how our solution works in practice, we give results on the construction of PSWF for the extended source shown in Fig. 3. The source was observed by the LOFAR ({http://www.lofar.org}) radio telescope during system testing. For an 8 hours of observation, we obtain, $N_a= 7896636$ data points on the Fourier plane. We reduce this to $N_a^{'}=11041$ points by applying our downsampling scheme as explained in section 3.1.

We use \cite{DT} to generate conforming Delaunay triangulations for the ROI and the exterior of the image. Selection of triangle scales for inside and outside of the ROI (see Fig.5), has to be done in such a way that the least amount of information is lost. For this observation, the PSF minor diameter is about $d=4.2\times 10^{-5}$ radians. To achieve the most appropriate triangulation, we have set different scales for the triangles inside the ROI ($a\times d$) and outside the ROI ($b\times d$), where $a,b$ are the parameters that we vary. 

We show the eigenvalue spectrum of the kernel $\mathbf{K}$, for different values of $a$ and $b$ in Fig. 6. The eigenvalue spectrum gives us a measure of information a certain basis can represent. By the results shown in Fig. 6, we see that in most settings, we get about $100$ basis functions that carry significant information, which is close to the limit given by Landau-Pollak theorem for this case. 

Once we obtain the basis functions for the downsampled grid, we interpolate back to the original image grid to get the basis vectors corresponding to the original image. Let ${\bf P}$ be the matrix whose columns correspond to the interpolated basis vectors. Then, we study the Frobenious norm of $\mathbf{P}^H\mathbf{P}-\mathbf{I}$ and the condition number of $\mathbf{P}^H\mathbf{P}$ in Fig. 7. This, in addition to the eigenvalue spectrum in Fig. 6 give us a benchmark to choose the most suitable triangulation (or values of $a$ and $b$). If the basis vectors are orthonormal, we will obtain the lowest value for the Frobenious norm. Fig. 7 (a) shows that the optimality of the source modeling is not affected significantly from one setting to another, because the basis vectors are orthonormal with negligible errors. Thus, the CRB remains minimum \cite{SAM2010}. Moreover, the condition number of $\mathbf{P}^H\mathbf{P}$ helps us to understand how its condition affects the further computation. Ultimately, these conclude that the setting with $a= 0.2$ and $b = 2.1$ is the most suitable one to fulfill the aforementioned criteria. By this setting, we downsample to $N^{'}=3270$ pixels in which the ROI has $N_b^{'}=3199$ pixels. In contrast, the original pixel grid has $N=7552$ and $N_b=1826$.

\begin{figure}[htbp]
\begin{minipage}{0.98\linewidth}
\centering
 \centerline{\epsfig{figure=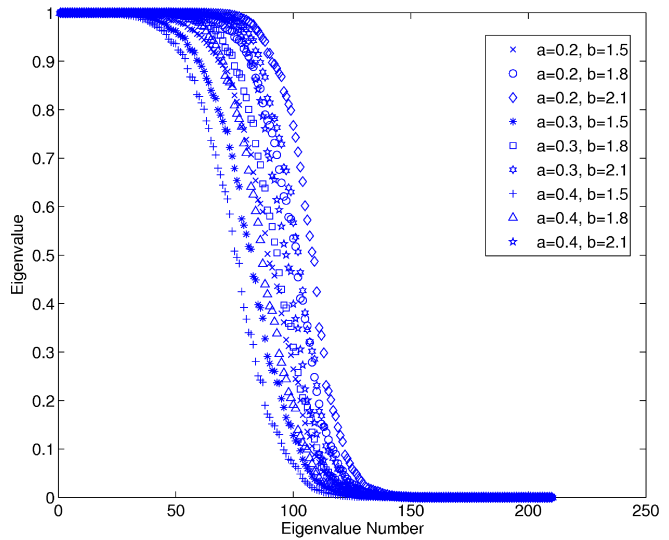,width=9.0cm}}
\end{minipage}
\caption{Eigenvalue spectrums for different triangulations, with varying $a$ and $b$.\label{eigs}}
\end{figure}

%\begin{figure}[htbp]
%\begin{minipage}{0.98\linewidth}
%\centering
% \centerline{\epsfig{figure=CondFrobNorm_a&b_Indexes.eps,width=9.0cm}}
%\end{minipage}
%\caption{(a) Condition number for different triangulations (varying $a$ and $b$) of the image. (b) Frobenious norm (scaled by $1/N_{basis}^2$ where $N_{basis}$ is the number of basis functions) for different %triangulations.\label{condn}}
%\end{figure}

\begin{figure*}[htbp]
\centering
\includegraphics [width=14.0cm] {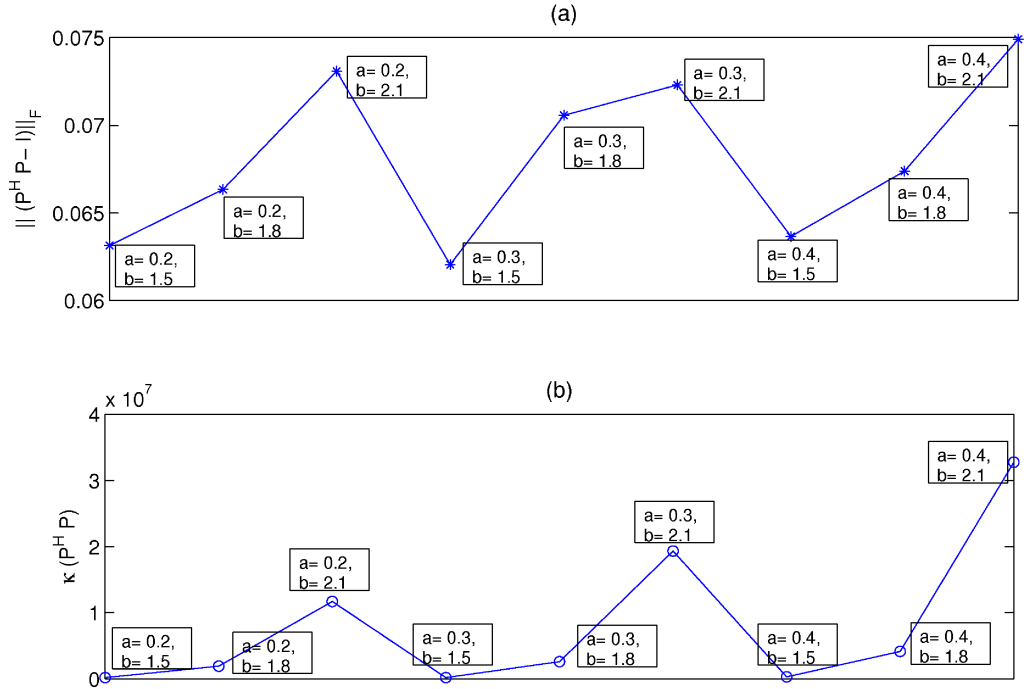}
\caption{(a) Frobenious norm of $\mathbf{P}^H\mathbf{P}-\mathbf{I}$ (scaled by $1/N_{basis}^2$ where $N_{basis}$ is the number of basis functions) for different triangulations (varying $a$ and $b$) of the image. (b) Condition number of $\mathbf{P}^H\mathbf{P}$ for different triangulations. \label{condn}}
\end{figure*}
%\begin{figure}[htbp]
%\begin{minipage}{0.98\linewidth}
%\begin{minipage}{0.98\linewidth}
%\centering
% \centerline{\epsfig{figure=Cond_ab_Paper.eps,width=9.00cm}}
%\vspace{0.1cm} \centerline{(a)}\smallskip
%\end{minipage}\\
%\begin{minipage}{0.98\linewidth}
%\centering
% \centerline{\epsfig{figure=FrobNorm_ab_Paper.eps,width=9.00cm}}
%\vspace{0.1cm} \centerline{(b)}\smallskip
%\end{minipage}
%\end{minipage}
%\caption{(a) Condition number for different triangulations (varying $a$ and $b$) of the image. (b) Frobenious norm (scaled by $1/N_{basis}^2$ where $N_{basis}$ is the number of basis functions) for different triangulations.\label{condn}}
%\end{figure}

\section{Conclusions}
The use of PSWF for source modeling and deconvolution of extended sources 
was proposed earlier in \cite{SYURSI2011} and \cite{ICIP2011}.
In this paper, we suggested a scheme to reduce its computational
cost. By using the Delaunay triangulation, we can significantly reduce the cost of construction of PSWF. We have also shown by a real example that we do not lose information by this cost reduction. Further improvement will focus on finer selection of the ROI depending on the source structure such that an area with higher intensity is represented with more triangle pixels. 

\bibliographystyle{IEEEtran}
\bibliography{IEEEpswf}

\end{document}